\documentclass[twocolumn]{aastex62}
\usepackage{lineno}

\newcommand{\kms}{km~s$^{-1}$}

\begin{document}

\title{Young and eccentric: the quadruple system HD 86588}

\author{Andrei Tokovinin}
\affiliation{Cerro Tololo Inter-American Observatory, Casilla 603, La Serena, Chile}
\author{Hank Corbett}
\affiliation{Department of Physics and Astronomy, University of North Carolina at Chapel Hill, Chapel Hill, NC 27599-3255, USA}
\author{Octavi Fors}
\altaffiliation{Dept. de Física Qu\`antica i Astrof\'{i}sica, Institut de Ci\`encies del Cosmos (ICCUB),
Universitat de Barcelona, IEEC-UB, Mart\'{\i} i Franqu\`es 1, E08028 Barcelona, Spain}
\affiliation{Department of Physics and Astronomy, University of North Carolina at Chapel Hill, Chapel Hill, NC 27599-3255, USA}
\author{Ward Howard}
\affiliation{Department of Physics and Astronomy, University of North Carolina at Chapel Hill, Chapel Hill, NC 27599-3255, USA}
\author{Nicholas M. Law} 
\affiliation{Department of Physics and Astronomy, University of North Carolina at Chapel Hill, Chapel Hill, NC 27599-3255, USA}
\author{Maxwell Moe} 
\affiliation{Steward Observatory, University of Arizona, 933 N. Cherry Ave., Tucson, AZ 85721, USA}
\author{Jeffrey  Ratzloff} 
\affiliation{Department of Physics and Astronomy, University of North Carolina at Chapel Hill, Chapel Hill, NC 27599-3255, USA}
\author{Frederick M. Walter}
\affiliation{Department of Physics and Astronomy, Stony Brook University, Stony Brook, NY 11794-3800, USA}

\email{atokovinin@ctio.noao.edu}
\correspondingauthor{Andrei Tokovinin}

\begin{abstract}
High-resolution  spectroscopy and  speckle  interferometry reveal  the
young star  HD 86588  as a quadruple  system with a  3-tier hierarchy.
The 0\farcs3 resolved  binary A,B with an estimated  period around 300
years contains  the 8-year pair Aa,Abc  (also potentially resolvable),
where Ab,Ac is a double-lined  binary with equal components, for which
we  compute the  spectroscopic  orbit.  Despite  the  short period  of
2.4058 day, the orbit  of Ab,Ac is eccentric ($e=0.086\pm0.003$).  It
has a  large inclination, but there  are no eclipses; only  a 4.4 mmag
light  modulation    apparently   caused  by  star  spots  on  the
  components of  this binary is detected with  Evryscope.  Assuming a
moderate extinction of $A_V = 0.5$  mag and a parallax of 5.2\,mas, we
find that  the stars are on or close to the  main sequence (age
$>$10 Myr) and  their masses are from 1 to 1.3  solar.  We measure the
strength of the Lithium line in the visual secondary B which, together
with rotation, suggests that the system is younger than 150 Myr.  This
object is located behind the  extension of the Chamaeleon I dark cloud
(which  explains extinction and  interstellar Sodium  absorption), but
apparently does  not belong  to it.  We  propose a scenario  where the
inner  orbit  has  recently  acquired its  high  eccentricity  through
dynamical interaction with   the outer two components;  it is now
undergoing rapid tidal circularization on a time scale of $\sim$1 Myr.
Alternatively, the  eccentricity could be  excited quasi-stationary by
the outer component  Aa.
\end{abstract} 


\section{Introduction}
\label{sec:intro}

The object  of this  study is a  young, chromospherically  active star
HD~86588 (Table~\ref{tab:par}).  \citet{Alcala1995} classified it as a
weak-lined T Tau star because of  the strong Lithium line; the star is
not known  to be variable.   Several authors \citep[e.g.][]{Frink1998}
attribute it  to the  Chamaeleon I (Cha  I) star forming  region.  The
{\it Gaia}  DR2 distance, 118\,pc, and  the corresponding heliocentric
velocity $(U,V,W) =  (+5.4, -3.6, -3.2)$ km~s$^{-1}$ (the  $U$ axis is
directed away from the Galactic center) are indeed similar to those of
the Cha~I members, although HD~86588 is located on the sky outside the
boundaries of  known molecular clouds.  No  infrared excess indicative
of  a debris  disc was  found by  \citet{Spangler2001}.  There  are no
emission lines in the spectrum.  \citet{Lopez2013}  attribute this
  star to the field by its kinematics.

\citet{Covino1997}  detected triple  lines in  the spectrum,  but they
have not  followed to determine  the spectroscopic orbit.   Later, the
same  team  found this  star  to be  ``single'',  based  on three  spectra
\citep{Guenther2007}.    Meanwhile,  it  was   resolved  in   1996  by
\citet{Koehler2001} into a 0\farcs27  visual pair denoted as KOH~86 in
the WDS \citep{WDS}. This object clearly deserves further study.

\begin{deluxetable}{l c  }
\tabletypesize{\scriptsize}     
\tablecaption{Main parameters of HD~86588
\label{tab:par} }  
\tablewidth{0pt}                                   
\tablehead{                                                                     
\colhead{Parameter} & 
\colhead{Value} 
}
\startdata
Identifiers & WDS J09527$-$7933,  RX J0952.7$-$7933 \\
Position (J2000) &  	$09^h 53^m 13\fs7365$ $-79^\circ 33' 28\farcs465$ \\
PM\tablenotemark{a} (mas~yr$^{-1}$) &  $-$11.8,  4.0 \\
Parallax\tablenotemark{b} (mas) & 8.50 $\pm$ 0.77 \\
Spectral type & F6V \\
$V$ (mag) & 9.63 \\
$K$ (mag) & 7.99
\enddata
\tablenotetext{a}{The PM is from {\it Tycho-2} \citep{Tycho2}.}
\tablenotetext{b}{The parallax is
  from the {\it Gaia} DR2 \citep{Gaia}.}
\end{deluxetable}

\begin{figure}
\epsscale{1.0}
\plotone{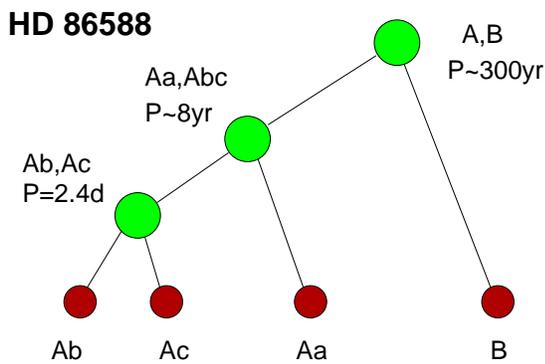}
\caption{Structure of the hierarchical quadruple system
  HD~86588. Green circles denote systems, smaller red circles are
  individual stars. 
\label{fig:mobile}
}
\end{figure}

We  found that  this system  is  quadruple, rather  than triple.   Its
hierarchical  structure  is  shown  in  Figure~\ref{fig:mobile}.   The
brightest and most  massive star Aa is a rapid  rotator, so its lines,
blended  with other components,  were not  previously detected  in the
spectra. The star Aa is orbited by the spectroscopic binary Ab,Ac. The
visual secondary B is at the upper level of the 3-tier hierarchy.

\vspace*{1cm}

\section{Spectroscopy}

\subsection{Observational data}

High-resolution optical  spectra used here  were taken with the  1.5 m
telescope sited at the  Cerro Tololo Inter-American Observatory (CTIO)
in   Chile   and   operated   by  the   SMARTS   Consortium,\footnote{
  \url{http://www.astro.yale.edu/smarts/}}  using  the CHIRON  optical
echelle spectrograph \citep{CHIRON}.  Observations were conducted from
March to June 2018.  Most spectra are taken in the  slicer mode with a
resolution of  $R=80,000$ and  a signal  to noise ratio  of 25  to 40.
Thorium-Argon calibrations were recorded for each spectrum. 

\begin{figure}
\plotone{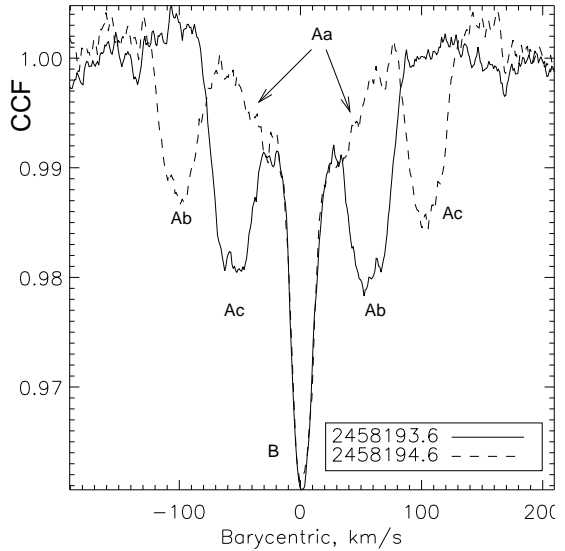}
\caption{CCFs of HD 86588 on two consecutive nights,
 JD 2459193.6 (full line) and  JD 2458194.6 (dashed line).
\label{fig:CCF}
}
\end{figure}

Radial  velocities (RVs)  were  measured by  cross-correlation of  the
reduced  spectrum  with the  binary  mask.   Further  details of  this
procedure can  be found in  \cite{paper1}.  Figure~\ref{fig:CCF} shows
the  cross-correlation  functions (CCFs)  recorded  on two  successive
nights.   The  strong  central  component  B  has  a  constant  RV  of
2\,km~s$^{-1}$,  while the  two  ``satellites'' Ab  and Ac  move
around it rapidly, indicating presence of the short-period subsystem.
\citet{Covino1997} measured  the RVs of  3 components at +2,  +85, and
$-98$ km~s$^{-1}$, in agreement with  these CCFs.  The fine details in
the CCFs are likely produced by spots on rapidly rotating stars in the
close  binary.  Naturally, these  stars are  chromospherically active,
explaining the X-ray detection.

\begin{figure}
\epsscale{1.15}
\plottwo{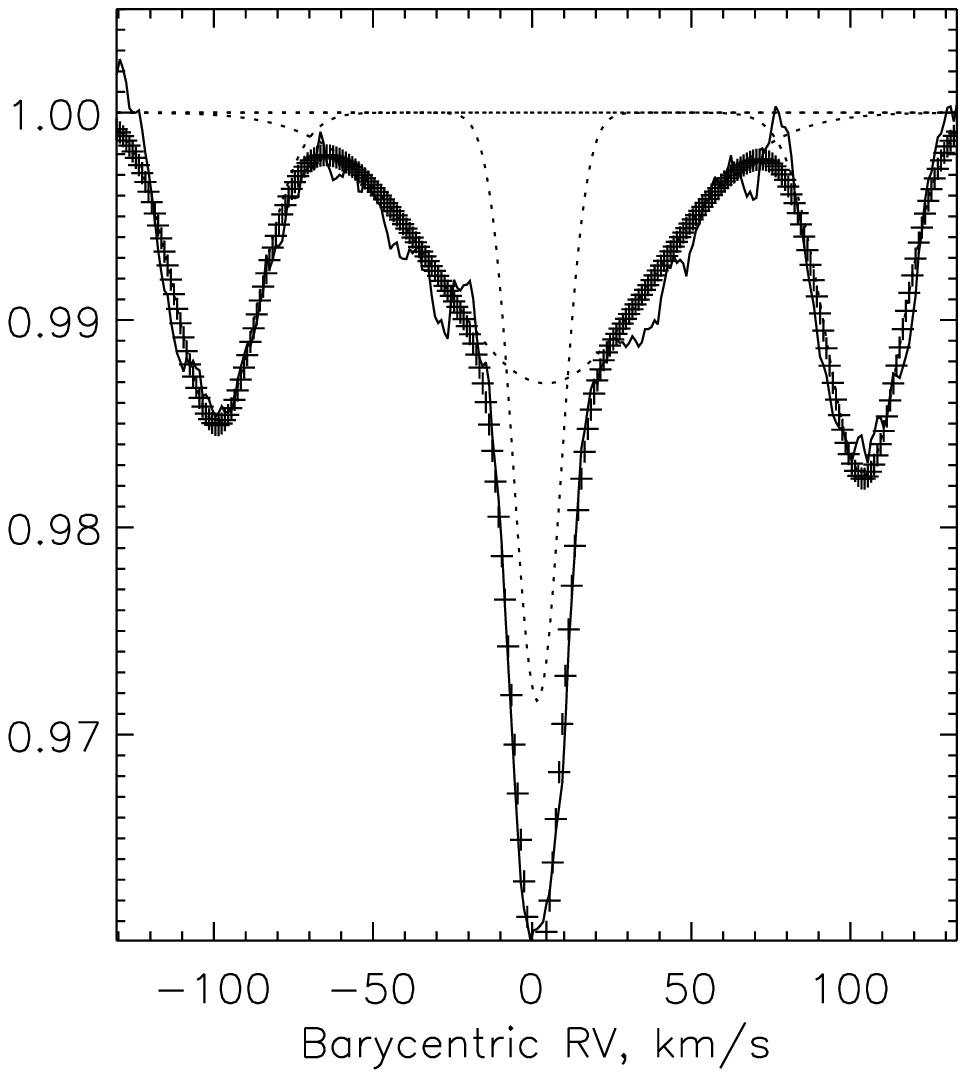}{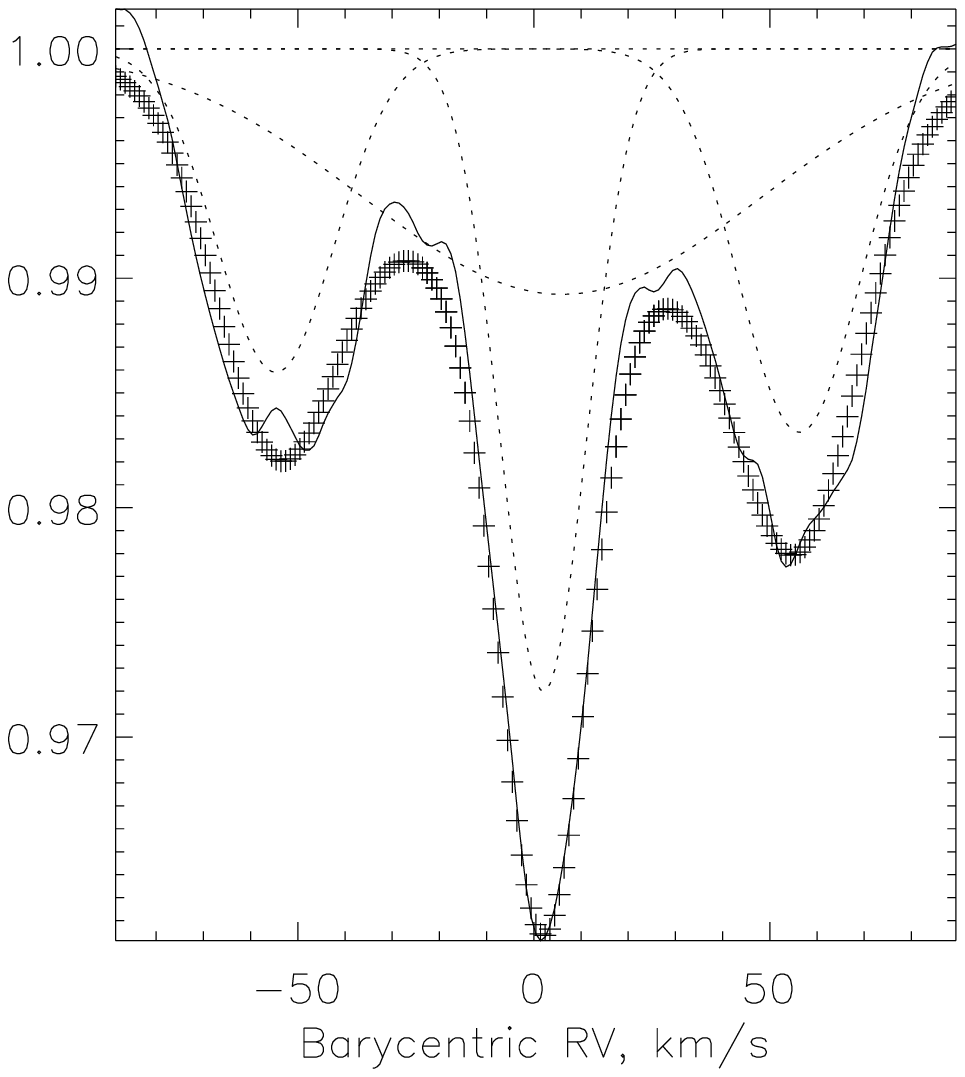}
\caption{Modeling of  the CCFs by 4  Gaussians. The full  curve is the
  CCF,  the pluses  are  the model,  and  the dotted  curves are  the
  individual  Gaussians.  Left:  JD  2458193.6, free  fit.  Right:  JD
  2458249.5, fitted with fixed parameters  of the broad component Aa. The RV
  of  Aa is  different from  the center-of-mass  RV of  Ab,Ac, causing
  asymmetry of the blended profile.
\label{fig:EW}
}
\end{figure}

In  most  resolved   triple-lined  multiple  systems,  the  stationary
component of the  CCF corresponds to one of  the resolved stars, while
the rapidly  moving lines  are produced by  the close  pair identified
with another star.  However, the  fluxes of the components B and Ab+Ac
are comparable,  while speckle photometry indicates that  B is fainter
than A by 2 mag.  One notes in Figure~\ref{fig:CCF} that the amplitude
of  the ``satellites''  is changing,  while the  central  component is
stable  in  both RV  and  amplitude.   When the  CCF  is  fitted by  3
Gaussians, the area of the components Ab and Ac depends on their RV in
a systematic  way: the satellites  become stronger when  they approach
the center and weaker when  they are widely separated.  Given the lack
of    strong    photometric    variability,   evidenced    below    in
Section~\ref{sec:var},  this  effect  cannot  be  caused  by  variable
circumbinary extinction.  Instead, it  is explained by blending of the
satellites with the fourth broad  CCF component Aa that can be guessed by
looking at the dashed line in Figure~\ref{fig:CCF}.

\begin{deluxetable}{l c c c  c }
\tabletypesize{\scriptsize}     
\tablecaption{Parameters of the CCF profile  derived from the 4-component fits
\label{tab:EW} }  
\tablewidth{0pt}                                   
\tablehead{                                                                     
\colhead{Parameter} & 
\colhead{Aa} &
\colhead{Ab} & 
\colhead{Ac} &
\colhead{B} 
}
\startdata
$a$                         & 0.0130 & 0.0155 & 0.0165 & 0.0298 \\
$\sigma$  (km~s$^{-1}$)     & 36.6   & 13.7 & 13.2     & 7.6 \\
$a \sigma$  (km~s$^{-1}$)    & 0.47 & 0.21 & 0.21 & 0.22 \\
RV  (km~s$^{-1}$)            & 7.8    & \ldots  & \ldots & 2.08 \\
$V \sin i$  (km~s$^{-1}$) &  66:   & 24:    &  23:  & 12.1 
\enddata
\end{deluxetable}

It turns out that the broad  central component of the CCF, Aa, has the
largest area and  corresponds to the brightest star  in this quadruple
system.   The 6 CCFs  with well-separated  satellites were  modeled by
fitting  4 Gaussians, and  average parameters  of each  component were
computed. The  remaining CCFs  where Aa is  heavily blended  with 
  the  other components  were fitted  by 4  Gaussians while   the
  amplitude, width,  and RV of the  broad component were  fixed to the
  average values. Figure~\ref{fig:EW} illustrates these  two cases.
One can  see in the  right panel that  blending with Aa  increases the
apparent depth of the satellites  and shifts their positions when they
are close to  each other.  In the 4-component  fits, the dependence of
the area of  components Ab and Ac on their position  (i.e.  on the RV)
vanishes.  The rms  residuals of the 4-component fits  to the CCFs are
typically about 0.001.

Table~\ref{tab:EW}  lists the mean  amplitude $a$  and the   mean
dispersion $\sigma$ of the Gaussians fitted to the subset of CCFs with
well-separated  dips. The product  $a \sigma$  is proportional  to the
area of  each component.   The rms scatter  of the amplitude  is about
0.001  and the  rms  scatter  of $\sigma$  is  about 0.4  km~s$^{-1}$.
Considering  the complex  nature of  the  CCF profiles  and the  small
amplitudes of  the dips,  the formal errors  of the  fitted parameters
should  be less  than their  real errors,  dominated by  the remaining
systematic distortions of the CCFs. For this reason, we do not provide
formal    errors   of    mean   parameters    in   Table~\ref{tab:EW}.
Table~\ref{tab:EW} also lists the average RVs of the two constant dips
Aa and  B and estimates of  the projected rotation  velocities $V \sin
i$,  computed  from  $\sigma$  by  the approximate  formula  given  in
\citep{paper1};   as this  formula is  valid for  solar-like stars
  with  $\sigma < 12$  km~s$^{-1}$, our  estimates of  $V \sin  i$ are
  crude, except for B. 

\begin{deluxetable}{l r r r r}    
\tabletypesize{\scriptsize}     
\tablecaption{Radial velocities
\label{tab:rv}          }
\tablewidth{0pt}                                   
\tablehead{                                                                     
\colhead{JD} & 
\colhead{Aa} &
\colhead{Ab} & 
\colhead{Ac} & 
\colhead{B} \\
\colhead{+2,400,000} &
\colhead{(km~s$^{-1}$)} &
\colhead{(km~s$^{-1}$)} &
\colhead{(km~s$^{-1}$)} &
\colhead{(km~s$^{-1}$)} 
}
\startdata
 58193.5983 & \ldots&  60.40 & $-$55.83 &  1.54  \\
 58194.5952 & 4.1   &$-$99.21 & 104.07  &  1.66  \\
 58195.6099 & 9.7   &  97.52 &$-$91.32  &  1.84  \\
 58228.5652 & \ldots&$-$38.73 &  42.81  &  2.09  \\
 58232.4992 & \ldots&$-$31.22 &  33.25 &   2.41  \\
 58242.4732 & 10.5  &$-$97.87 & 102.44 &   1.49  \\
 58246.4876 & 8.4   &  63.61 &$-$63.93 &   2.91  \\
 58248.6117 & 5.5   &  94.03 & $-$90.84 &  3.30  \\
 58249.4486 & \ldots&$-$55.43 &  57.41 &   1.93  \\
 58249.5358 & \ldots&$-$73.57 & 76.62 &    2.60  \\
 58249.6149 & \ldots&$-$87.88 &  91.58 &   2.90  \\
 58250.4750 & \ldots&  33.00 &$-$33.11 &   0.85  \\
 58250.5979 & \ldots&  58.37 &$-$57.69 &   1.89  \\
 58276.4479 & 8.4   &$-$93.89 &  94.82 &   2.68 
\enddata 
\end{deluxetable}



The individual RVs derived by  fitting CCF with 4 Gaussians are listed
in Table~\ref{tab:rv}.   The RVs of  the broad-lined component  Aa are
determined from the 6 CCFs  where the satellites are widely separated,
with large errors (rms scatter 2.5 \kms);  their mean value is 7.8
  km~s$^{-1}$.  The errors  for the other components are  on the order
  of 1 \kms.  The RVs of the central component B range from 0.9 to 3.3
  km~s$^{-1}$;  the  rms  scatter  of  0.8 km~s$^{-1}$  is  caused  by
  blending with other components.

\subsection{Spectroscopic orbit}

\begin{figure}
\plotone{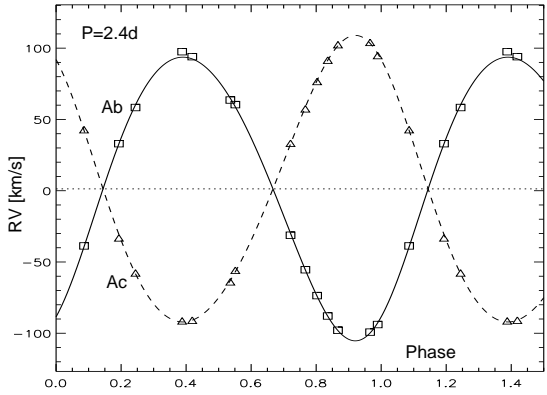}
\caption{Spectroscopic    orbit    of    the    close    pair    Ab,Ac 
  (see Table~\ref{tab:sb}).   Squares  and  full  line correspond  to  the
  component Ab, triangles and dashed line to Ac.
\label{fig:orb}
}
\end{figure}

As the  two satellites Ab  and Ac are  equal and move rapidly,  it was
difficult to tell  which is which in any  individual observation.  One
of us (F.W.) took three spectra  during one night and two more spectra
on the following night.  This  established the orbital period at about
2 days,  helping to tag the  components and to derive  the orbit.  The
elements   of  the  2.4   day  spectroscopic   orbit  are   listed  in
Table~\ref{tab:sb},  while Figure~\ref{fig:orb}  gives  the RV  curve.
Despite  the short  period, the  spectroscopic orbit  is significantly
non-circular.   This result, based  on the de-blended  RVs derived
  from the  4-component fits, also  holds when using the  original RVs
  derived  from  the  3-component  fits  that neglect  Aa.   The  rms
residuals to  the orbit  are 1.1  and 1.0 km~s$^{-1}$  for Ab  and Ac,
respectively  (the  residuals  were  2.4 and  1.6  km~s$^{-1}$  before
de-blending).  The  residuals are unusually large  for CHIRON spectra.
However, they are explained by  the complex and variable nature of the
CCF  profiles,  illustrated in  Figure~\ref{fig:CCF}.   If a  circular
orbit is enforced, the residuals increase to 5 km~s$^{-1}$.

\begin{deluxetable}{l c  }
\tabletypesize{\scriptsize}     
\tablecaption{Spectroscopic orbit of A{\rm b},A{\rm c}
\label{tab:sb} }  
\tablewidth{0pt}                                   
\tablehead{                                                                     
\colhead{Parameter} & 
\colhead{Value} 
}
\startdata
Period $P$ (day) &  2.4058 $\pm$  0.0002 \\
Periastron $T_0$ (JD) & 2458250.010  $\pm$          0.014 \\
Eccentricity $e$ & 0.086  $\pm$        0.003 \\
Longitude $\omega$ (\degr) &   213.8  $\pm$   2.1 \\
Primary amplitude $K_1$ (km~s$^{-1}$) &  99.50    $\pm$  0.51 \\
Secondary amplitude $K_2$ (km~s$^{-1}$) &  100.52    $\pm$  0.47 \\
$\gamma$ velocity  (km~s$^{-1}$) &  1.28     $\pm$  0.20 \\
R.M.S. residuals  (km~s$^{-1}$) &  1.09,     0.98 \\
$M \sin^3 i$ (${\cal M}_\odot$) & 0.990, 0.980 
\enddata
\end{deluxetable}

A star of one solar radius synchronized with the orbit would rotate at
20.9\,km~s$^{-1}$,  similar to our crudely estimated $V \sin i$.
This  fact, together with  the large  $M \sin^3  i$, implies  a highly
inclined orbit of Ab,Ac.

\subsection{The Lithium line}

\begin{figure}
\plotone{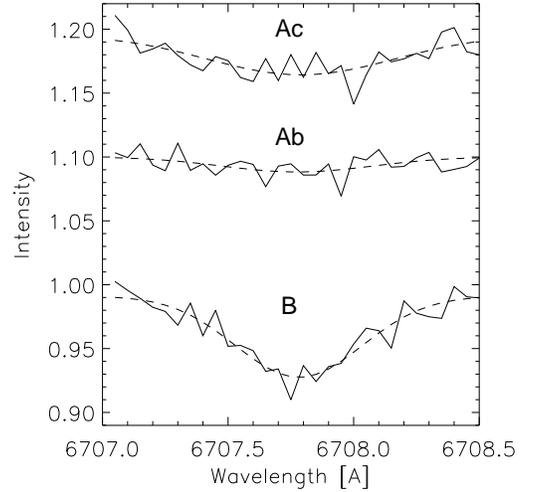}
\caption{Profiles  of the  lithium  6707.8\AA ~line  in the 
  components B, Ab, and Ac (from bottom up, displaced by 0.1 to avoid
  overlap). The dotted lines are fitted Gaussian curves.
\label{fig:Li}
}
\end{figure}

Knowing the  RVs of  each component, we  can extract  their individual
spectral  features by  modeling and  subtracting the  contributions of
other components, as explained  in \citep{paper1}.  Briefly, the depth
of the Lithium line in Ab, Ac,  and B (3 parameters, as we ignore here
the  line  of  Aa,  too  wide  to  be  measurable)  is  found  by  the
simultaneous  linear  fit  to  all  spectra  using  the  RVs  and  the
preliminary values  of the  line width. Then,  for one  component, the
lines  of  the  two  remaining  components are  subtracted  from  each
spectrum, all spectra are shifted  to the zero velocity, and co-added.
Figure~\ref{fig:Li} shows the Lithium line in three components derived
by this method.  It is clearly  present in B, with an equivalent width
of 40$\pm$5\,m\AA  (or $\sim$200\,m\AA  ~if corrected for  dilution by
other components   that contribute  0.8 of the light).   The wide
Lithium lines  in the satellites Ab  and Ac are  detectable, but their
equivalent  width  is  not  well  measured.  The  equivalent  width  of
200\,m\AA  ~reported   by  \citet{Alcala1995}  was   derived  from  the
low-resolution spectra where all components were blended. 

\subsection{Sodium lines}

\begin{figure}
\epsscale{1.1}
\plotone{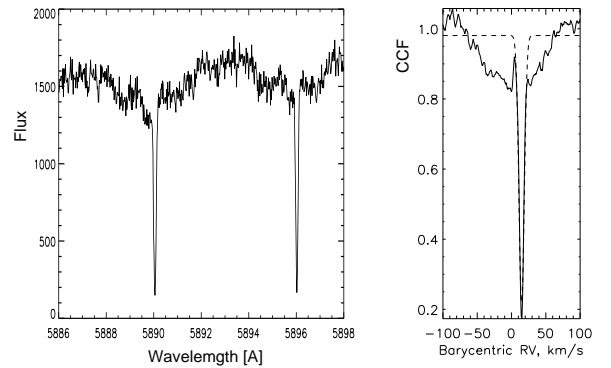}
\caption{Portion  of  the spectrum  around  the  Sodium  D lines  with
  interstellar  absorption.  The  spectrum  taken on  JD 2458193.6  is
  shown on the left (compare to the CCF in Figure~\ref{fig:CCF}).  The
  CCF with the  Sodium mask is plotted in the  right panel (the dashed
  line is  the fitted Gaussian) for  the spectrum taken  on JD 2458232.6,
  showing a weak emission on the left side of the interstellar dip. 
\label{fig:Na}
}
\end{figure}

Narrow  and deep  interstellar  lines  of Sodium  are  present in  the
spectra (Figure~\ref{fig:Na}).  The CCF  with a binary mask containing
only those two  lines shows that the RV of  these features is constant
at +14.5\,km~s$^{-1}$, while the  dispersion of the Gaussian fitted to
the CCF is only 3.4\,km~s$^{-1}$.   The narrow Sodium lines are caused
by material  on the line of  sight which is unrelated  to the multiple
system.  The RV  of the Sodium absorption matches the  RV of stars and
gas in  the Cha~I dark  cloud \citep{Covino1997}.  In addition  to the
absorption, some (but not all)  spectra have a weak emission component
in the D-lines (see the right panel of Figure~\ref{fig:Na}); the RV of
the emission is close to  zero.  The narrow Sodium emission could come
from the active chromosphere of the  component B or from the sky light
pollution.

\section{Variability}
\label{sec:var}

\begin{figure}
\plotone{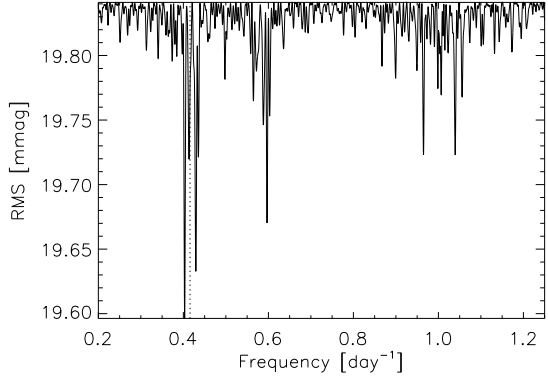}
\caption{Periodogram of  photometric variability.  The  plot shows the
  rms variance of the magnitude after subtracting sine and cosine terms fitted
  at each frequency. The vertical dotted line marks the orbital
  frequency.
\label{fig:ptm}
}
\end{figure}

Photometric variability  of HD~86588 is expected  for several reasons:
eclipses  in  the  close  binary,  ellipsoidal  variation,  star spots,
variable circumstellar  extinction.  Photometry in the  SDSS $g'$ band
was  provided  by  the  Evryscope  instrument sited  at  Cerro  Tololo
\citep{Evryscope}.  A total of 17640 measurements covering the period
from JD 2457755 to 2458078 with a typical cadence of 2.2 min yield the
mean magnitude of 8.503 mag (in uncalibrated instrumental system) and
the rms  scatter of  19.84 mmag.  Therefore,  to the first  order, the
star is not  variable on the time scales from minutes  to a year.  The
close binary  is not eclipsing.  Variable  circumstellar extinction is
also excluded.

Figure~\ref{fig:ptm} shows  the periodogram  in the period  range from
0.8 to 5 days (there is no significant variability at longer periods).
It is computed by fitting sine  and cosine terms at each trial period,
subtracting the fit, and estimating the variance of the residuals. The
strongest details  are located  near the frequency  of 0.4  cycles per
day;  the  orbital frequency   of  the  close binary is  0.4155
day$^{-1}$.   Subtraction  of the  largest  sine  term   with  the
  frequency of  0.403 day$^{-1}$ reduces the rms  from 19.84\,mmag to
19.60\,mmag, hence the rms variability at this frequency is 3.07\,mmag
and   the  corresponding  sine-wave   amplitude  is   4.4\,mmag.   The
micro-variability is apparently caused by star spots in the components
Ab and  Ac. They  are rotating  with periods that  are close,  but not
exactly equal to the binary period.

Components   of    slightly   eccentric   binaries    usually   rotate
pseudo-synchronously  with a frequency  $1 +  6e^2$ times  the orbital
frequency \citep[see equation  43 in][]{Hut1981}, or 0.4338 day$^{-1}$
for   the    Ab,Ac   binary.     The   two   strongest    details   in
Figure~\ref{fig:ptm} have  frequencies of 0.403  and 0.429 day$^{-1}$.
If they correspond to the  rotation periods of the two components, one
of them can  be almost pseudo-synchronized, while the  other rotates a
bit slower.   The tidal pseudo-synchronization timescale  is orders of
magnitude faster  than the  tidal circularization timescale   (see
  Section~\ref{sec:origin}),   so    it   is   consistent    with   a
pseudo-synchronized  binary still having  a slightly  eccentric orbit.
In such a case, the X-ray  emission  would be driven  by tidal spin up  as the
system is currently experiencing rapid tidal circularization.

The  variability at  the frequency  of  0.6 day$^{-1}$,  also seen  in
Figure~\ref{fig:ptm}, could be caused by star spots on another rapidly
rotating star in this system,  either Aa or B. 

Assuming the mass  sum of 2 ${\cal M}_\odot$  for Ab,Ac, the semimajor
axis of the close binary is  0.044 au or 9.5 $R_\odot$. The absence of
eclipses  restricts the  inclination  of the  spectroscopic binary  to
$i_{\rm Ab,Ac} <  78^\circ$, or $\sin^3 i <  0.935$. The minimum masses
of the stars Ab and Ac are, therefore, 1.06 ${\cal M}_\odot$.

\section{Positional measurements}

Apart  from  the discovery  measure  in  1996 by  \citet{Koehler2001},
observations  of the  relative motion  of  the outer  binary A,B  
  (where A refers to the photo-center of the unresolved inner system)
come  from two  sources.   \citet{Vogt2012} monitored  this pair  with
adaptive optics at  VLT from 2007 to 2011.   Furthermore, the relative
position  was  measured  by  speckle interferometry  at  the  Southern
Astrophysical  Research  Telescope  (SOAR)  from  2011  to  2018  
  \citep[see][and references therein]{SAM18}.   Most speckle measures
at SOAR  were made in  the $I$ band  and result in the  mean magnitude
difference $\Delta  I_{\rm AB}  = 1.80$ mag,  with the rms  scatter of
0.12 mag.  Additionally,  we measured $\Delta V_{\rm AB}  = 1.90$ mag.
The  totality of measures  provides a  relatively dense  coverage from
2007 to 2018.

\begin{figure}
\plotone{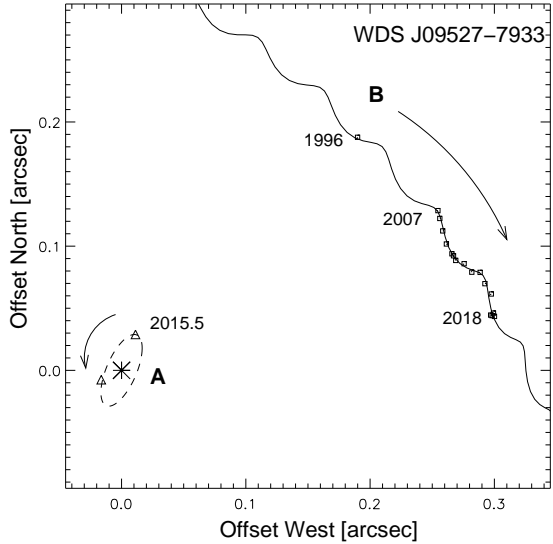}
\caption{Motion of  the visual  companion B around  A, located  at the
  coordinate  center. Scale in  arcseconds, North  up, East  left. The
  wavy line  is discussed in the  text.  The dashed  ellipse denotes a
  possible orbit  of Aa,Abc, the triangles depict   its positions 
    in 2015.5 and 2018.0.
\label{fig:pos}
}
\end{figure}

The pair A,B is in slow retrograde motion. It turned by 36\degr ~in 20
years  since its  discovery in  1996.  This  motion corresponds  to an
orbital period of  a few hundred years. The  period estimated from the
projected separation is  of the same order of  magnitude.  However, as
shown  in   Figure~\ref{fig:pos},  the  observed   motion  is  ``wavy'',
suggesting presence  of unresolved subsystem with a  period of $\sim$8
years. This additional system presumably corresponds to the Aa,Abc
pair, as shown in the diagram in Figure~\ref{fig:mobile}.

\begin{deluxetable}{l c c  }
\tabletypesize{\scriptsize}     
\tablecaption{Tentative visual orbits
\label{tab:vis} }  
\tablewidth{0pt}                                   
\tablehead{                                                                     
\colhead{Parameter} & 
\colhead{A,B} &
\colhead{Aa,Abc}
}
\startdata
$P$ (year)    & 300 & 8.0 \\
$T_0$ (year)  & 1865.0 & 2023.4 \\
$e$         & 0.07    &  0 \\
$a$ (\arcsec)& 0.43 & 0.0066 \\
$\Omega$ (degree) & 44.0 & 336.6 \\
$\omega$ (degree) & 289.0   & 0 \\
$i$ (degree) & 126.2 & 67.0 
\enddata
\end{deluxetable}

Table~\ref{tab:vis} gives two sets  of orbital elements describing the
observed motion of A,B.   The outer orbit is not constrained by the
short observed  arc. It is chosen  to represent the data  and to match
the  mass  sum  and  parallax  from  the  system  model  developed  in
Section~\ref{sec:model}.    The   circular   inner   orbit      in
  Table~\ref{tab:vis}  is sufficient  to  model the  wobble, but  the
actual inner orbit can be  eccentric just as well.  These elements fit
the data, but are by no  means unique; they are one possible solution.
The rms residuals to this model are 1.3\,mas in both coordinates; they
increase to  4\,mas if  the wobble is  ignored.  The  measurements and
residuals     to    the    tentative     orbits    are     given    in
Table~\ref{tab:speckle}.  Note  that the  systems A,B and  Aa,Abc have
opposite directions  of orbital motion, meaning that  their orbits are
definitely not coplanar.

\begin{deluxetable}{ l rr rr l}    
\tabletypesize{\scriptsize}     
\tablecaption{Position measurements and residuals
\label{tab:speckle}          }
\tablewidth{0pt}                                   
\tablehead{                                                                     
\colhead{Date} & 
\colhead{$\theta$} & 
\colhead{$\rho$} & 
\colhead{(O$-$C)$_\theta$ } & 
\colhead{(O$-$C)$_\rho$ } &
\colhead{Ref.\tablenotemark{a}} \\
\colhead{(year)} &
\colhead{(\degr)} &
\colhead{(\arcsec)} &
\colhead{(\degr)} &
\colhead{(\arcsec)} &
}
\startdata
 1996.2480  & 314.6 & 0.2670   &    0.1 &   0.0013 & K \\
 2007.1626  & 296.9 & 0.2852   &   $-$0.1 & 0.0005 & V \\
 2008.1320  & 295.6 & 0.2838   &    0.1 &$-$0.0006 & V \\
 2009.1368  & 293.6 & 0.2818   &    0.2 &$-$0.0007 & V \\
 2010.1469  & 291.4 & 0.2805   &    0.3 &$-$0.0007 & V \\
 2010.9773  & 289.6 & 0.2819   &    0.3 &$-$0.0002 & V \\
 2011.2284  & 289.2 & 0.2826   &    0.4 &$-$0.0004 & S \\
 2011.0370  & 288.4 & 0.2831   &   $-$0.8 & 0.0007 & S \\
 2012.1020  & 287.5 & 0.2887   &    0.1 & 0.0018 & S \\
 2013.1288  & 285.9 & 0.2927   &   $-$0.4 &$-$0.0003 & S \\
 2014.0436  & 285.5 & 0.2991   &   $-$0.1 & 0.0012 & S \\
 2015.9135  & 283.7 & 0.3006   &      0.0 &$-$0.0013 & S \\
 2016.9595  & 281.9 & 0.3038   &    0.2 & 0.0025 & S \\
 2018.0868  & 279.0 & 0.3028   &      0.1 & 0.0012 & S \\
 2018.4012  & 278.7 & 0.3014   &    0.1 &$-$0.0008 & S \\
 2018.4012  & 278.5 & 0.3032   &   $-$0.1 & 0.0010 & S \\
 2018.4012  & 278.8 & 0.3003   &    0.2 &$-$0.0019 & S 
\enddata 
\tablenotetext{a}{
K: \citet{Koehler2001};
V: VLT \citep{Vogt2012};
S: speckle interferometry at SOAR.
}
\end{deluxetable}

The period of 8 years, mass sum of 3.4\,${\cal M}_\odot$, and parallax
of  5.2  mas   (see  Section~\ref{sec:model})  correspond to  the
semimajor axis of 6 au or 31\,mas on the sky.  The wobble amplitude of
6.6\,mas  is only 21\%  of the  semimajor axis,  being reduced  by the
comparable  fluxes and  masses of  Aa and  Abc.  The  subsystem Aa,Abc
should  be resolvable  by speckle  interferometry.  It  was definitely
unresolved at SOAR in 2018.0  and 2018.4, while all other observations
do not reach the full diffraction-limited resolution  of SOAR, 40\,mas.

The center-of-mass velocity of  Ab,Ac  and the velocity of Aa should vary
in anti-phase with  a period of $\sim$8 years and  an amplitude of the
order of $\sim$10\,km~s$^{-1}$. At  present, the RV difference between
Aa and Abc (the  center of mass of the inner binary) is  $ 7.8 - 1.3 =
6.5$   \,km~s$^{-1}$.    The   RVs   of   Ab  and   Ac   measured   by
\citet{Covino1997}   imply    that   Abc    had   an   RV    of   $(85
-98)/2=-6.5$\,km~s$^{-1}$,   different  from   its  actual   value  of
1.3\,km~s$^{-1}$.  Hence,  the 8 year orbit does  produce the expected
RV    signature.    \citet{Guenther2007}    measured    the   RV    of
2.4\,km~s$^{-1}$  and considered the  star to  be single;  most likely,
their RVs refer to the narrow-lined component B. Apparently, its RV is
constant.

The short-term proper motion (PM)  measured by  {\it Gaia} DR2 \citep{Gaia},
$(-4.0, -3.1)$ mas~yr$^{-1}$, differs from the long-term {\it Tycho-2}
PM reported in Table~\ref{tab:par}.   If the latter corresponds to the
center-of-mass motion,  the differential PM in 2015.5  (the DR2 epoch)
was  10.5 mas~yr$^{-1}$, directed  at 133\degr  ~angle.  The  orbit of
Aa,Abc  in  Table~\ref{tab:vis} predicts  motion  of the  photo-center
between 2015 and 2016 with  a speed of 11.7\,mas~yr$^{-1}$ directed at
77\degr.   The PM  difference  between {\it  Gaia}  and {\it  Tycho-2}
roughly matches the proposed orbit in speed, but not in direction.  We
predict that {\it Gaia} will soon detect astrometric acceleration and,
possibly, will derive the astrometric orbit of Aa,Abc.

\section{Modeling}
\label{sec:model}

\begin{deluxetable}{l c cc c  c }
\tabletypesize{\scriptsize}     
\tablecaption{Magnitudes and masses
\label{tab:model} }  
\tablewidth{0pt}                                   
\tablehead{                                                                     
\colhead{Parameter} & 
\colhead{Aa} &
\colhead{Ab} &
\colhead{Ac} &
\colhead{B} &
\colhead{A} 
}
\startdata
$V$ (mag) & 10.37 & 11.53 & 11.53 & 11.70 & 9.80 \\
$K$ (mag) & 8.80 & 9.96 & 9.96 & 9.73 & 8.23 \\
${\cal M}$ (${\cal M}_\odot$) & 1.30 & 1.05 & 1.05 & 1.02 & 3.40
\enddata
\end{deluxetable}

We  make the  first, preliminary  attempt  at modeling  the system  by
evaluating individual magnitudes of  the components and comparing them
to the  isochrones. The photometry  by \citet{Vogt2012} yields  a good
measurement of $\Delta m_{\rm AB} = 1.50 \pm 0.12$ mag at 2.18\,$\mu$m
wavelength that can be identified  with $\Delta K_{\rm AB}$. They also
measured  $\Delta m_{\rm AB}  = 1.55$  mag at  1.265\,$\mu$m.  Speckle
interferometry at SOAR gives $\Delta I_{\rm AB}$ = 1.8 mag and $\Delta
V_{\rm AB} = 1.9$ mag   with errors of $\sim$0.1 mag.  This, 
  together  with the  total  brightness of  the  system, defines  the
individual $V$ and $K$ magnitudes of the resolved components A and B.

The  areas of  the CCF  dips, corrected  slightly for  the temperature
dependence by  using preliminary assumed  effective temperatures, lead
to the  flux fractions of 0.54, 0.16,  0.16, and 0.14 for  Aa, Ab, Ac,
and B, respectively, in the $V$ band. The corresponding $\Delta V_{\rm
  AB} = 1.9$ mag matches  the speckle photometry.  The relative fluxes
of the  unresolved components Aa, Ab, and  Ac in the $K$  band are not
measured. They are computed by  assuming $\Delta K \approx 0.65 \Delta
V$,  as  derived from standard relations for  main sequence stars
of these  masses; the measured  $V$-band flux ratios are  re-scaled to
the $K$ band. This defines the individual magnitudes of all components
listed in the first two lines of Table~\ref{tab:model}. 

\begin{figure}
\plotone{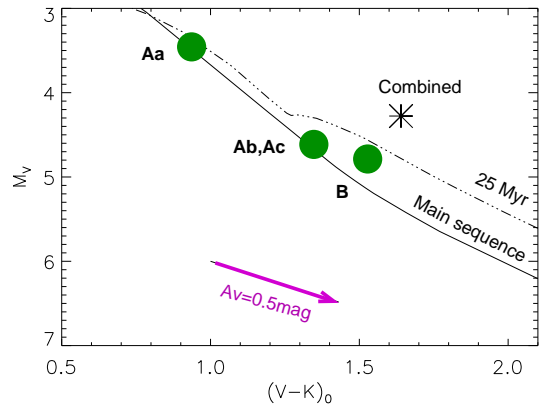}
\caption{Location  of the  components (filled  circles) on  the  1 Gyr
  isochrone (line) assuming parallax of  5.2\,mas and $A_V = 0.5$ mag.
  The  magenta arrow is  the reddening  vector. The  big asterisk
  corresponds to  the combined light  and the {\it Gaia}  DR2 parallax
  without extinction.  The dash-dot  line is the 25-Myr PARSEC isochrone 
  \citep{PARSEC}.
\label{fig:cmd}
}
\end{figure}

We  begin by  comparing these  stars  to the  1 Gyr  solar-metallicity
Dartmouth isochrone  \citep{Dotter2008}.  The {\it  Gaia} DR2 parallax
of  8.50$\pm$0.77 mas places  the components  slightly below  the main
sequence.   The corresponding  masses of  Ab  and Ac  are 0.85  ${\cal
  M}_\odot$, contradicting their minimum mass of 1.06 ${\cal M}_\odot$
derived from the  orbit. However, the DR2 astrometry  of this star has
unusually  large errors,  likely caused  by the  duplicity and  by the
acceleration  in the 8-year  orbit.  Poor  quality of  the astrometric
solution for  HD~86588 is manifested  by the excess noise  of 5.2\,mas
(its typical value in DR2 is 0.2 mas). Therefore, the DR2 parallax can
be biased. 

The total extinction in the direction  of HD~86588 is $A_V = 0.48$ mag
according          to         \citet{dust}.\footnote{See         \url{
    https://irsa.ipac.caltech.edu/applications/DUST/}}  The  agreement
between model and observations is improved if we adopt $A_V = 0.5$ mag
and a smaller parallax of  5.2\,mas. This makes the stars brighter and
more  massive (circles  in  Figure~\ref{fig:cmd}).  The  main-sequence
masses of  all components,  also given in  Table~\ref{tab:model}, then
roughly  match the minimum  masses of  Ab and  Ac.  The  extinction is
confirmed   by   the    presence   of   interstellar  lines   in
Figure~\ref{fig:Na}.  The RV of these lines indicates that the dust is
related to the Cha~I dark cloud, while the RV of HD~86588 is different
by $\sim$10\,\kms.

We  stress  that  the  proposed  system  model  is  based  on  several
assumptions and is  by no means definitive. The $V-K$  color of Ab and
Ac is not  measured directly, hence their location  on the main sequence
is artificial. The  measured color of the component  B places it above
the main sequence,  under the 25-Myr isochrone.

The age of the system can be inferred from the strength of the Lithium
6708\,\AA ~line.   The component B  contributes about 0.2  fraction of
the total  flux at this  wavelength, so the measured  equivalent width
(EW) of  40\,m\AA ~translates  to the intrinsic  EW of  0.2\,\AA.  The
effective temperature  inferred from  the isochrone is  5920\,K ($\log
T_e   =  3.77$,  spectral   type  G0V).   According  to   Figure~6  of
\citet{Covino1997}, the star is  located  above the upper envelope
for  the   Pleiades,  somewhere   among  massive  members   of  Cha~I.
Apparently,  the system is  younger than  the Pleiades  (i.e.  younger
than $\sim$150\,Myr), but we cannot say how much younger.

Stellar rotation  provides another  diagnostic of age.   The projected
rotation  velocity of  the solar-mass  component B,  12.2 km~s$^{-1}$,
implies  its rotation period  $P_{\rm rot,  B} <  4.1$ day.   The fast
rotation of Aa, estimated  very crudely from its CCF, corresponds
to $P_{\rm rot,  Aa} < 1$ day.  Adopting the $B-V$  colors of 0.63 and
0.43  mag  for  B and  Aa,  respectively,    derived  from  the
  isochrones, the  age-rotation relation of  \citet{Barnes2007} leads
to the minimum  ages of 150 and 100 Myr. If  the periodogram detail at
0.6 day$^{-1}$ corresponds to the  true rotation period of B, the same
relation gives the age of 25\,Myr.

\section{Origin of the eccentric inner binary}
\label{sec:origin}

Non-zero eccentricity  of the inner binary Ab,Ac  provides an estimate
of its ``tidal''  age, i.e.  the time of  tidal orbit circularization.
Using the  equilibrium-tide model with convective damping  and a scale
factor   of   $F_{\rm   tid,con}    =   20$,   as   parameterized   in
\citet{Belczynski2008},  a pair of  stars with  $M_1 =  M_2 =  1 {\cal
  M}_\odot$, $R_1 =  R_2 = 1 R_\odot$ and a period  of 2.4 day evolves
from $e = 0.3$ to $e=0.1$  in 0.8 Myr.  At the pre-main sequence (PMS)
stage, when the stellar radii  are larger, the tidal evolution is much
faster;  PMS  binaries with  periods  shorter  than  $\sim$6 days  are
already  circularized  \citep{Meibom2005}.  So,  the  inner binary  in
HD~86588 has  a tidal  age of $\sim$1~Myr,  much less than  the actual
system age.   Somehow, the  binary had to  become eccentric  after its
components reached the main sequence.

As suggested by \citet{Moe2018},  most eccentric binaries with periods
below   the  circularization   period   acquired  their   eccentricity
relatively  recently  through  dynamical interaction  in  hierarchical
systems (Kozai-Lidov, or K-L, cycles).  Therefore, their ``tidal age''
is a  small fraction  of the true  age.  However,  compact hierarchies
with outer periods  as short as 8 years tend  to have nearly co-planar
orbits, preventing the  K-L cycles. Moreover, even if  the orbits were
inclined, the period of K-L  cycles  and the associated time scale
  of orbit  evolution are much  shorter than the system  age, leaving
the paradox of young tidal age unsolved.

The fact that  this system is quadruple, rather  than triple, comes to
the rescue.  \citet{Hamers2015} studied dynamics of quadruple
systems with 3+1 hierarchy composed of ``nested'' binaries X, Y, and Z
(in our case,  the 2.4-day, 8-year and 300-year  systems).  They found
that the  inner binary X can  acquire a large eccentricity  even if it
were originally coplanar with the intermediate binary Y.  This happens
when  the periods  of K-L  cycles in  the inner  triple X-Y  and outer
triple Y-Z  are comparable.   The K-L period  for the inner  triple is
given by their equation (11), namely
\begin{equation}
P_{\rm KL, X-Y} = \frac{P_{\rm Y}^2}{P_{\rm X}} \; \frac{M_{\rm X1} + M_{\rm X2} +
  M_{\rm Y2}}{M_{\rm Y2}} (1 - e_{\rm Y}^2)^{3/2} .
\label{eq:PKL} 
\end{equation}
The K-L  period in the outer triple  Y-Z is computed in  the same way.

Using  the  parameters  determined  above, we  estimate  for  HD~86588
$P_{\rm KL, X-Y} \approx 26$ kyr  and $P_{\rm KL, Y-Z} \approx 47$
kyr. So, these  periods are comparable and the  dynamical evolution of
the inner  binary to  large eccentricity by  this mechanism  is 
possible, at least in principle. 

An  alternative mechanism,  frequently invoked  in the  literature, is
constant excitation  of the inner  binary's eccentricity by  the outer
companion (in our  case Aa), accompanied by tidal  damping.  The small
eccentricity  is  then  defined  by  the  balance  of  these  opposite
processes and persists for a long time. The long-lived eccentric inner
binary makes  its discovery more  probable, compared to the  chance of
catching  it   during  the  fast   tidal  damping  episode.    If  the
eccentricity  of  the  Aa,Abc orbit    is $e_{\rm Y}  <  0.8$,  then  general
relativistic  (GR)  and  tidal  precession currently  suppresses  such
continuous excitation of the inner binary's eccentricity
\citep{Liu2015}.  However,  this  orbit likely  oscillates
between  small  and  large  eccentricities  due  to  K-L  cycles  with
component  B.   The  K-L  timescale  $P_{\rm KL, Y-Z} =  47$  kyr  is  also
substantially  shorter  than the  tidal  circularization timescale  of
$e/\dot{e} \sim 1$ Myr.  If the 8-year orbit can reach $e_{\rm Y} > 0.8$ in its
K-L cycle,  then it can sustain  the eccentricity $e_{\rm X} =  0.09$ of the
inner binary,  counteracting tidal  friction as well  as GR  and tidal
precession.

\section{Summary and discussion}
\label{sec:disc}

Originally, the  interest in HD~86588  was driven by its  presumed PMS
status related to its membership in Cha~I.  We found that this star is
located  behind  the  Cha~I  molecular cloud,  which  imprints  sodium
absorption in  its spectrum  and causes a  mild extinction. The  RV of
HD~86588  differs from the  RV of  the Cha~I  group by  $\sim$10 \kms.
However, this  system is  definitely juvenile (younger  than $\sim$150
Myr), as  evidenced by  the presence of  strong lithium line  and fast
rotation.  It  could be born in  a previous episode  of star formation
related to Cha~I.

Complex nature of this  3-tier hierarchical system adds uncertainty to
its interpretation.  We cannot trust the {\it Gaia} parallax until the
8-year wobble is  accounted for in its astrometric  solution. Our best
guess  is a  parallax  of 5.2\,mas  (distance  190\,pc); assuming  the
extinction of $A_V = 0.5$ mag,  we propose a model where the stars Aa,
Ab, and Ac are located on the main sequence and their masses match the
minimum  mass  derived  from  the spectroscopic  orbit.   The  adopted
parallax and the system velocity of 3.5 \kms (estimated center-of-mass
RV of  Aa,Abc) lead to the  revised heliocentric motion  of $(U,V,W) =
(+8.7,  -6.1, -5.3)$  \kms. This  velocity  does not  match any  known
kinematic group.

The close binary Ab,Ac has  a non-zero eccentricity $e= 0.09$, unusual
for its short period of 2.4\,day.  However, this binary is not unique.
The  multiple  star catalog  \citep{MSC}  contains many  spectroscopic
binaries belonging to hierarchical  systems. Among 184 such pairs with
periods less than 4 days and primary mass less than 1.5 solar, 30 have
non-zero eccentricity.  Eight of them  have a complex hierarchy with 3
or 4 tiers, reminiscent of HD~86588,  while the rest are triple or 2+2
quadruple systems  with just 2 tiers.    An example  of the second
  group  is   MSC  16228$-$2326,  a  PMS  2+2   quadruple  system  RX
J1622.7$-$2325  where  the  pair  Ba,Bb  composed of  two  0.4  ${\cal
  M}_\odot$   stars   has   $P   =   3.23$  day   and   $e   =   0.30$
\citep{Rosero2011}.  These  authors show the  period-eccentricity plot 
where the PMS and main  sequence binaries occupy the same locus.  They
argue  that eccentric  orbits with  short  periods are  found in  both
groups,  independently of age,  and suggest  that eccentricity  can be
excited  by dynamical  interaction with  other components  in multiple
systems.

Although  HD~86588 is  no longer  interesting as  a calibrator  of PMS
evolution,   its  further   study  can   clarify    mechanisms  of
close-binary  formation within stellar  hierarchies. The  next obvious
step would be a spatial  resolution of the 8-year subsystem, either by
speckle  interferometry  at  8  m  telescopes,  or  by  long  baseline
interferometers  such as  VLTI. In  parallel,  spectroscopic monitoring
during  several years and  the {\it  Gaia} astrometry  will complement
interferometry in  defining the intermediate orbit.  This will provide
accurate measurement of stellar masses  for all components.

\acknowledgements

We thank the operator of the 1.5-m telescope R.~Hinohosa for executing
CHIRON  observations of  this  program and  L.~Paredes for  scheduling
observations.  Re-opening  of CHIRON  in 2017 was  largely due  to the
enthusiasm and  energy of  T.~Henry.  The Evryscope  team acknowledges
funding support by the  National Science Foundation grants ATI/1407589
and CAREER/1555175. O.F.  acknowledges  funding support by the Spanish
Ministerio  de Econom\'ia  y Competitividad  (MINECO/FEDER,  UE) under
grants  AYA2016-76012-C3-1-P,   MDM-2014-0369  of  ICCUB   (Unidad  de
Excelencia  ``Mar\'ia de  Maeztu'').    Detailed  comments by  the
  Referee helped to clarify the presentation.

This work  used the  SIMBAD service operated  by Centre  des Donn\'ees
Stellaires  (Strasbourg, France),  bibliographic  references from  the
Astrophysics Data  System maintained  by SAO/NASA, and  the Washington
Double Star Catalog maintained at USNO.

\facilities{CTIO:1.5m, SOAR, Evryscope}










\end{document}